\documentclass[journal]{IEEEtran}
\usepackage{cite}

\ifCLASSINFOpdf
   \usepackage[pdftex]{graphicx}
   \DeclareGraphicsExtensions{.pdf,.jpeg,.png}
\else
   \usepackage[dvips]{graphicx}
   \DeclareGraphicsExtensions{.eps}
\fi
\usepackage[cmex10]{amsmath}
\usepackage{algorithmic}

\usepackage{array}

\usepackage{mdwmath}
\usepackage{mdwtab}

\usepackage{eqparbox}

\usepackage[tight,footnotesize]{subfigure}

\usepackage{balance}
\usepackage[ansinew]{inputenc}
\usepackage{times} 
\usepackage{tikz}
\usepackage[none]{hyphenat}
\usetikzlibrary{shapes,arrows}

\graphicspath{{figures/}}

\begin{document}
%
\tikzstyle{inicio} = [rectangle,draw,minimum height=1em, text
width=5em, text centered]

\tikzstyle{bloque} = [rectangle, draw,
    text width=5em, text centered, minimum height=1em, text
width=10em]

\tikzstyle{bloque1} = [rectangle, draw,
    text width=5em, text centered,  minimum height=1em, text
width=5em]

\tikzstyle{linea} =[draw,-latex, very thin]

\tikzstyle{decision} = [diamond, draw,
    text width=7em, text badly centered, node distance=2.5cm, inner sep=0pt, minimum height=1em]

\tikzstyle{line} = [draw, -latex', very thin]
\title{Band-phase-randomized Surrogates to assess nonlinearity in non-stationary time series}
%
%
%

\author{Diego~Guarín$^{*}$,~\IEEEmembership{Student~Member,~IEEE,}
        Edilson~Delgado,~\IEEEmembership{Member,~IEEE,}
        and~Álvaro~Orozco.
\thanks{D. Guarin and A. Orozco are with the Department
of Electrical Engineering, Universidad Tecnológica de Pereira - UTP,
Vereda la Julita, Pereira, Colombia.}
\thanks{E. Delgado is with the Research Center of the Instituto Tecnológico Metropolitano – ITM, Calle 73 No. 76A-354 Vía al volador, Medellín, Colombia.}
\thanks{$^{*}$Corresponding author. email: dlguarin@ieee.com}}
%
%

\markboth{IEEE Transactions on Biomedical Engineering,~Vol.~?, No.~?, January~2011}%
{D. Guarín \MakeLowercase{\textit{et al.}}: Bare Demo of
IEEEtran.cls for Journals}
%



\maketitle

\begin{abstract}
Testing for nonlinearity is one of the most important preprocessing
steps in nonlinear time series analysis. Typically, this is done by
means of the linear surrogate data methods. But it is a known fact
that the validity of the results heavily depends on the stationarity
of the time series. Since most physiological signals are
non-stationary, it is easy to falsely detect nonlinearity using the
linear surrogate data methods. In this document, we propose a
methodology to extend the procedure for generating constrained
surrogate time series in order to assess nonlinearity in
non-stationary data. The method is based on the
band-phase-randomized surrogates, which consists (contrary to the
linear surrogate data methods) in randomizing only a portion of the
Fourier phases in the high frequency band. Analysis of simulated
time series showed that in comparison to the linear surrogate data
method, our method is able to discriminate between linear
stationarity, linear non-stationary and nonlinear time series. When
applying our methodology to heart rate variability (HRV) time series
that present spikes and other kinds of nonstationarities, we where
able to obtain surrogate time series that look like the data and
preserves linear correlations, something that is not possible to do
with the existing surrogate data methods.
\end{abstract}

%
\IEEEpeerreviewmaketitle
\section{Introduction}
%
%
%
%
\IEEEPARstart{T}{}he surrogate data method, initially introduced by
J. Theiler et al. \cite{theiler} is nowadays one of the most popular
tests used in nonlinear time series analysis to investigate the
existence of nonlinear dynamics underlying experimental data. The
approach is to formulate a null hypothesis for a specific process
class and compare the system output to this hypothesis. The
surrogate data method can be undertaken in two different ways:
\emph{Typical realizations} are Monte Carlo generated surrogates
from a linear model that provides a good fit to the data;
\emph{constrained realizations} are surrogates generated from the
time series to fulfill the null hypothesis and to conform to certain
properties of the data. The latter approach is suitable for
hypothesis testing due to the fact that it does not requiere pivotal
statistics \cite{theiler2}. In order to test a null hypothesis at a
certain confidence level, one has to generate a given number of
surrogates. Then, one evokes whatever statistic is of interest and
compares the value of this statistic computed from the data to the
distribution of values elicited from the surrogates. If the
statistic value of the data deviates from that of the surrogates,
then the null hypothesis may be rejected. Otherwise, it may not.\\
The linear methods for constrained realizations namely (i) Random
shuffle (RS); (ii) Random phase (RP); and, (iii) Amplitude adjusted
Fourier transform (AAFT) surrogates \cite{theiler}, were developed
to test the null hypothesis that the data came from a (i) i.i.d
gaussian random process, (ii)  linear correlated stochastic process;
and (iii) nonlinear static transformation of a linear stochastic
process. Surrogates generated with the RS method are constrained to
the amplitude distribution ($\mathcal{AD}$) or rank distribution of
the original data, while the ones generated with the RP algorithm
preserve the autocorrelation ($\mathcal{AC}(\tau)$) and surrogates
generated with the AAFT algorithm preserve both the $\mathcal{AD}$
and $\mathcal{AC}(\tau)$ of the original data.\\
As the process that generates surrogate data is stationary
\cite{Kaplan1997}, there could be some situations where surrogates
fail to match the data, even though the $\mathcal{AD}$ and
$\mathcal{AC}(\tau)$ are the same for the data and surrogates, so
the null hypothesis could be trivially rejected. This is particulary
true when data are non stationary. Because of this, when the
statistical properties of data are time dependent it is not feasible
to use the linear surrogate data methods for testing nonlinearity
\cite{Schreiber2000} (Timmer \cite{Timmer1998} showed that for some
non-stationary processes the test is able to discriminate between
linear and nonlinear data, but
this is not a general result). \\
From the introduction of the linear surrogate data method, there has
been a widespread interest in modifying it to assess nonlinearity in
non-stationary time series. The first attempt (as we can tell) to
apply the method to non-stationary time series was done by T.
Schreiber \cite{Schreiber1998}. He proposed that to deal with
non-stationarity data, the null hypothesis should include it
explicitly. Because otherwise, the rejection of a null hypothesis
can be equally to nonlinearity or non-stationarity. e.g., given any
process we can ask whether the data is compatible with the null
hypothesis of a correlated linear stochastic process with time
dependent local behavior. In order to answer this question in a
statistical sense we have to create surrogate time series that show
the same linear correlations and the same time dependency of the
local behavior as the data and compare a nonlinear statistic between
data and surrogates \cite{Schreiber2000}. To generate surrogates
constrained to data $\mathcal{AC}(\tau)$ and time dependence of
local behavior, T. Schreiber  \cite{Schreiber1998} used an iterative
procedure called simulated annealing. Unfortunately, this method
requires a big amount of
computational time and never became of popular usage.\\
In another study, A. Schmitz and T. Schreiber
\cite{Schmitz99surrogatedata} proposed a different method to deal
with non-stationarity. The proposed method involved dividing the
signal into stationary segments, then applying the linear surrogate
data method to each segment and finally joining the segments to form
a surrogate time series of the same size as the original data. The
major problem with this procedure is that there is not a
straightforward way to find stationary segments in a
non-stationary signal.\\
Recently, T. Nakamura and M. Small \cite{tomomichi} proposed a new
methodology to apply the surrogate data method to time series with
trends, called Small Shuffle Surrogate (SSS) data method which is a
modification of the RS algorithm. The main idea introduced in
\cite{tomomichi} is that in order to preserve the trend of the data
in surrogates, the randomization should be applied only in a small
scale, in this way all local correlations in the original time
series are destroyed in surrogates; but the global behavior (i.e.,
the trend) is preserved.\\ Based on the idea of preserving the slow
behavior of the signal in surrogates, T. Nakamura et al.
\cite{tomomichi2} presented a modification of the RP algorithm which
makes it suitable for data with trends. They called it the Truncated
Fourier Transform Surrogate (TFTS) data method. TFTSs are
constrained to conform to the $\mathcal{AC}(\tau)$ and with the
correct parameter selection to the trend of data (the authors also
apply the modification to the iAAFT method, thus preserving the
$\mathcal{AD}$, $\mathcal{AC}(\tau)$ and the trend of data in
surrogates). So, nonstationarities (in this case caused by the
presence of a trend) are included in the null hypothesis, as
suggested by A. Schmitz and T. Schreiber \cite{Schreiber1998,
Schreiber2000}. The idea of the method is to preserve the slow
behavior or trends while destroying all possible nonlinear
correlations in the irregular fluctuations. To achieve this goal,
the authors proposed to randomize phases only in the
higher-frequency domain and not alter the low-frequency phases (the
original idea of band-phase-randomized surrogates was briefly
proposed by J. Theiler et al. \cite{Theiler1993} but it was not
implemented until the work of T. Nakamura et al. \cite{tomomichi2}).
This approach is in contrast to linear surrogate methods (RP and
iAAFT), where all phases are randomized.
\\
It is worth mentioning that other attempts have been made in order
to assess nonlinearity in non-stationay data. L. Faes et al.
\cite{Faes2009-2} presented a method for calculating the parameters
of an non-stationary AR model. Based on this method, they generated
typical realizations of the non-stationary Heart Rate Variability
(HRV) signals and tested for nonlinearity, but as the surrogates are
typical realizations, one needs a pivotal statistic. Recently, C.J.
Keylock \cite{Keylock2007} presented a modification of the iAAFT
method based on the wavelet transform, with this method it is
possible to generate surrogates constrained to preserve the
$\mathcal{AC}(\tau)$ and the local mean and variance of the data,
but according to our personal experience the method proposed by T.
Nakamura et al. \cite{tomomichi2} is simpler to implement and
achieves similar results. In a recent publication \cite{Guarin2010},
we presented a modification of the TFTS through which we assessed
nonlinearity in data with spikes, but this method is limited to data
with gaussian $\mathcal{AD}$.\\
In this document we introduce the band-phase-randomized surrogate
methods in a rather organized way, we also present the algorithms to
facilitate and promote the application of the method. In regards to
the method, we present a discussion on the parameter selection and
introduce some modifications to the parameter selection criteria in
order to make the method suitable for different types of
nonstationarities (not only trends). To test the proposed
methodology we applied the test to several simulated time series
with different dynamical properties. We also applied the methods to
HRV signals of healthy patients. Finally we conclude.
\section{Background}
Prior to introducing the current technology in surrogate data
methods, it is vital to make one observation: Hypothesis testing,
such as the surrogate data method, cannot be used to determine what
the data \emph{is}, only what the data \emph{is not}
\cite{theiler2}. That is; if after our comparison we cannot
distinguish between data and surrogates, this may be simply because
our selected statistic is inadequate. Conversely, if the data and
surrogates are different, then we can sate, that, with a certain
probability the data is not consistent with the corresponding null
hypothesis.
\subsection{Surrogate data methods}
\subsubsection{Linear surrogate data methods} Linear surrogate data
were introduced to preclude a linear filtered noise source as the
possible origin of experimental data. The algorithms, as stated
earlier, generate surrogate data that fulfill the null hypothesis of
IID noise; linearly filtered noise; and, a monotonic nonlinear
transformation of linearly filtered noise. Hence, these techniques
produce flawless linear data. The algorithms to generate such
surrogates can be stated as follows \cite{theiler}.
\begin{description}
  \item[\textbf{RS}] A surrogate time series $\{s_{t}\}$ is
  generated from the scalar time series data $\{x_{t}\}$ by randomly
  shuffling $\{x_{t}\}$. This process destroys all temporal
  correlations (which are not expected in a IID process) but
  maintains the same $\mathcal{AD}$.
  \item[\textbf{RP}] The surrogate $\{s_{t}\}$ is generated by
  taking the Fourier transform of the data, randomising the phases
  (replacing it by the phases of a random IID process of the same
  length as $\{x_{t}\}$), and taking the inverse Fourier transform.
  The surrogate therefore maintains the linear correlations of data
  but any nonlinear structure is destroyed.
  \item[\textbf{AAFT}] One first re-scales the data
  original time series so that it is Gaussian, then generates an
  Algorithm 1 surrogate of the data $\{p_{t}\}$, and finally
  re-orders the original data so that it has the same rank
  distribution as $\{p_{t}\}$. This re-ordered time series
  constitutes the surrogate $\{s_{t}\}$. This process achieves two
  aims: first, just as with the Algorithm 1, the power spectra (and
  therefore linear correlations) of data is preserved in surrogates;
  second, the re-ordering process means that the $\mathcal{AD}$ of
  data and surrogates are also identical.
\end{description}
It should be noted that the AAFT algorithm does not deliver what it
promises. The phase randomisation will preserve the linear
correlation, but re-scaling the output of the inverse Fourier
transform $\{p_{t}\}$ to have the same $\mathcal{AD}$ as the
original data will alter the autocorrelation structure of the data.
Although the data and surrogate will have identical rank
distribution, the linear correlation will only be approximately the
same. A solution to this problem has been proposed by T. Schreiber
and A. Schimitz \cite{Schreiber1996}. Essentially, the solution is
to iterate the AAFT algorithm until convergence is achieved.
However, there is no guarantee that this iteration will, in fact,
converge. This algorithm is refereed to as improved AAFT (for a
discussion on the convergence of the iAAFT algorithm see
\cite{Venema2006}). \subsubsection{Surrogate data methods for data
with trends} 
\begin{figure}
 \centering
\begin{tikzpicture}[scale=2, node distance = 1.3cm, auto]
  \node [inicio] (te2) {Time Series};
  \node [bloque1, below left of=te2, node distance=2cm] (te4) {Magnitude};
  \node [bloque1,  below right of= te2, node distance=2cm] (te5) {Phase};
  \node [bloque, below of=te5, node distance=2cm] (te6) {Randomize the phases in a portion of the higher frequency domain};
  \node [bloque, below of=te2, node distance=6.2cm] (te8) {Truncated Fourier Transform Surrogates};
  \path [line] (te2) -- node {FT} ++(0,-0.35) -- +(-0.7,0) -- (te4);
  \path [line] (te2) -- ++(0,-0.35) -- +(0.7,0) -- (te5);
  \path [line] (te5) -- (te6);
  \path [line] (te4) -- ++(0,-1.9) -- +(0.7,0) -- (te8);
  \path [line] (te6) -- ++(0,-0.9) -- +(-0.71,0) --  node {IFT} (te8);
\end{tikzpicture}
  \caption{Flow chart of the Truncated Fourier Transform Surrogate method.}\label{fig2}
\end{figure}
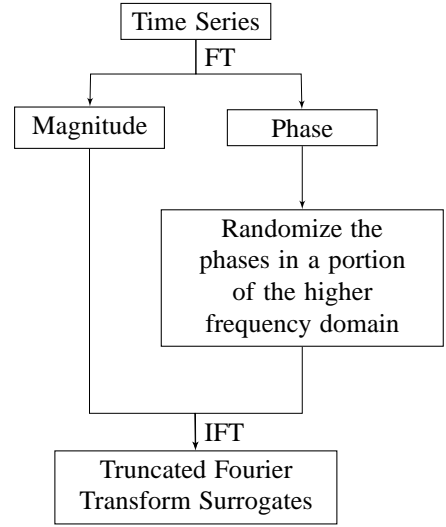

As stated earlier, when data are non-stationary, the
hypothesis addressed by the linear surrogate data methods are
trivially rejected. Two different surrogate data methods have been
proposed to tackle data with trends, the SSS and the TFTS data
methods. The hypothesis tested by SSS algorithm is that the data,
while possibly exhibiting some trend, is otherwise just noise
\cite{tomomichi}; while the hypothesis tested by TFTS algorithm is
that the data, while possibly exhibiting some trend, is generated by
a stationary linear system \cite{tomomichi2}. These algorithms can
be stated as follow \cite{small6}.
\begin{description}
  \item[\textbf{SSS}] Let $\{i_{t}\}$ be the index of $\{x_{t}\}$ (that is, $i_{t}=t$ and so
  $x_{i_{t}}=x_{t}$). Obtain $\{i'_{t}\}=\{i_{t}+Ag_{t}\}$ where
  $\{g_{t}\}$ are Gaussian random numbers, and $A$ is an amplitude (note that $\{i_{t}\}$ will be a sequence
  of integres, whereas $\{i'_{t}\}$ will not). Rank order $\{i_{t}\}$
  to obtain $\{r_{t}\}$. The surrogates $\{s_{t}\}$ are obtained from
  $s_{t}=x_{r_{t}}$. If $A$ is an intermediate value (e.g. 1),
  surrogates generated by this algorithm will preserve the slow
  trend
  in the data, but any inter-point dynamics will be destroyed by the local shuffling of individual points.
  \item[\textbf{TFTS}] The surrogate $\{s_{t}\}$ is generated by
  taking the Fourier transform of the data
  $\{X_{\omega}\}_{\omega}$. Then generating random phases
  $\phi_{\omega}$, such that $\phi_{\omega} \sim U(0,2\pi)$ if $\omega >
  f_{c}$ and $0$ if $\omega \leq f_{c}$ ( $\phi_{\omega}$ have to be antisymmetric around $\phi_{0}$). Finally taking the inverse
  Fourier transform of the complex series $\{X_{\omega}e^{\imath \phi_{\omega}}\}_{\omega}$ (Fig.
\ref{fig2}). As in the RP surrogates, all linear dependencies are
preserved in surrogates. But, since some phases are untouched, TFTS
data may still have nonlinear correlations. However, it is possible
to discriminate between linear and nonlinear data because the
superposition principle is only valid for linear data, so when data
are nonlinear, even if the power spectrum is preserved completely,
the inverse Fourier transform data using randomized phases will
exhibit a different dynamical behavior
\end{description}
TFTSs are influenced primarily by the choice of frequency $f_{c}$.
If $f_{c}$  is too high, surrogates are almost identical to the
original data. In this case, even if there is nonlinearity in the
data, one may fail to detect it. Conversely, if $f_{c}$ is too low,
surrogates are almost the same as the linear surrogate and the local
behavior is not preserved. In this case, even if there is no
nonlinearity in the data, one may wrongly judge otherwise.\\
In general, the correct value of $f_{c}$ cannot be determined
\emph{a priori}. To select an adequate value of $f_{c}$, T. Nakamura
et al. \cite{tomomichi2} proposed to start randomizing a portion of
the higher frequency domain (e.g. a 1$\%$ of the higher frequency
domain, i.e., $f_{c} \approx N/2$), decreasing $f_{c}$ until the
data linearity is no longer preserved in the surrogates (i.e.,
$\mathcal{AC}(\tau = 1$) of data falls outside the distribution of
surrogates) and then perform the test with the last value of $f_{c}$
for which linearities of data are preserved in surrogates.
\subsection{Significance and power of the test}
Applying a statistical hypothesis test to observed data can result
in two outcomes:  either the null hypothesis is rejected, or it is
not. In the former case there is a probability $\alpha$ that the
null hypothesis is rejected even though it is true (Type I Error),
in the latter case there is a probability $\beta$ (Type II Error)
that we will fail to reject the null when it is in fact false.  The
probability $\alpha$ is known as the significance level, its
complement ($1 - \alpha$) is the confidence level. For example, if
one generates 19 surrogates using some algorithm, and these yield a
larger (or smaller) value of some statistic than the data, then the
probability that this result occurred by chance is
$\alpha=\frac{1}{20}$, and hence we conclude at the 0.05
significance (0.95 confidence) level for a one-sided test that the
selected statistic is different from the surrogates.  Conversely,
the power of a test ($1-\beta$) is the probability the null
hypothesis is correctly rejected.\\
Clearly, the probability $\alpha$ is determined by the number of
trials and the number of independent test statistics. Computing
$\alpha$ is only a matter of computing probabilities. The problem is
that the value of $\beta$ is not clear.  The actual power $\beta$
will depend on the choice of test statistic.  If the test statistic
is independent of data and surrogates then the power is determined
by the number of trials \cite{small6}.
\section{Nonlinearity test for non-stationary time series: Physiological data approach}
\subsection{Database}
\subsubsection{Simulated time series} To test the proposed
methodology we applied it to different simulated time series, two
linear (stationary and non-stationary) and two nonlinear (stationary
and non-stationary). The linear time series were generated by the
following AR(2) process \cite{Timmer1998}
\begin{equation}\label{ec1}
    x(n)=a_{1}(n)x(n-1)+a_{2}x(n-2)+\eta.
\end{equation}
Where
\begin{equation}\label{ecu2}
\begin{split}
    a_{1}(n)&=2cos(2\pi/T(n))e^{(-1/\tau)}, \ a_{2}=e^{(-2/\tau)},\\
    T(n)&= T_{e}+M_{T}sin(2\pi t/T_{mod}),\\
    \eta& \sim \mathcal{N}(0,1).
\end{split}
\end{equation}
To generate a linear stationary signal we used $T_{e}=10$,
$T_{mod}=250$, $\tau=50$ and $M_{T}=0$, for the linear
non-stationary signal we used $M_{T}=6$.\\ The nonlinear time series
were generated by the following nonlinear process \cite{Faes2009-1}
\begin{equation}\label{ec3}
    x(n)=a_{1}(n)x(n-1)(1-x^{2}(n-1))e^{(-x^{2}(n-1))}+a_{2}x(n-2).
\end{equation}
For the nonlinear stationary signal we used $a_{1}(n)=3.4$ and
$a_{2}=0.8$. For the nonlinear non-stationary signal we used
\begin{equation*}
    a_{1}(n)=
    \begin{cases}
    3.0 & \text{if $0 < n \leq N/2$,}\\
    3.4 & \text{if $N/2 < n \leq N$.}
    \end{cases}
\end{equation*}
\begin{figure}
\centering
  \includegraphics[width=9cm]{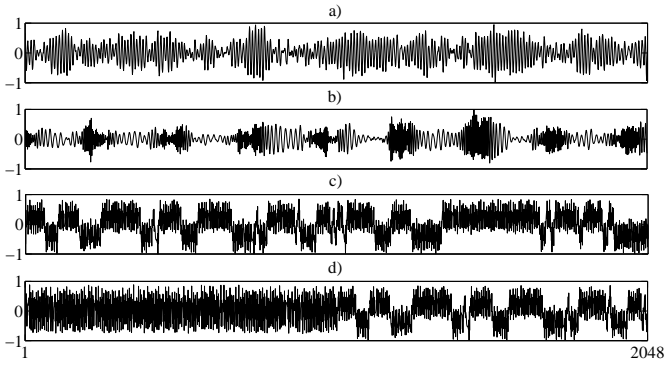}\\
  \caption{a) Linear stationary (LS) signal, b) linear
non-stationary (LNS) signal, c) nonlinear stationary (NLS) signal
and d) nonlinear non-stationary (NLNS) signal. }\label{fig1}
\end{figure}
An example of each of these signals is shown in Fig. \ref{fig1} with
$N=2048$. \subsubsection{Physiological time series} The HRV time
series of healthy subjects were extracted from  the MIT-BIH  Normal
Sinus Rhythm Database in Physionet \cite{Glass2009, Goldberger2000}
according to annotations for only normal beats. Sample  rate  was
128  Hz  in 24-hr  Holter recordings.
\subsection{Proposed procedure}
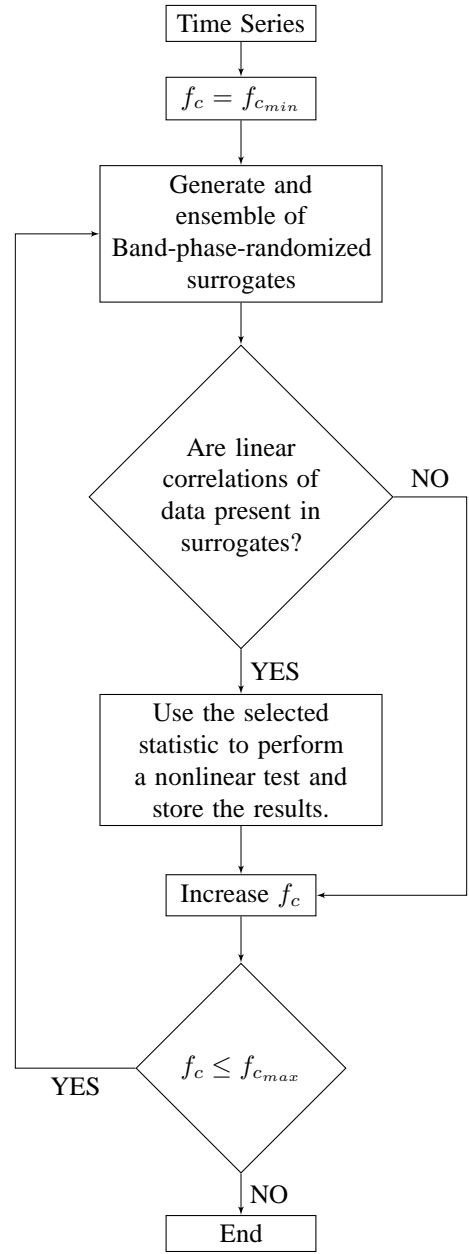
\begin{figure}
 \centering
\begin{tikzpicture}[scale=2, node distance = 1cm, auto]
  \node [inicio] (te1) {Time Series};
  \node [bloque1, below of=te1] (te2) {$f_{c}=f_{c_{min}}$};
  \node [bloque,  below of=te2, node distance=1.8cm] (te3) {Generate and ensemble of Band-phase-randomized surrogates};
  \node [decision, below of=te3, node distance=3.5cm] (te4) {Are linear correlations of data present in surrogates?};
  \node [bloque, below of=te4, node distance=3.5cm] (te5) {Use the selected statistic to perform a nonlinear test and store the results.};
  \node [bloque1, below of=te5, node distance=1.8cm] (te6) {Increase $f_{c}$};
  \node [decision, below of=te6, node distance=2.3cm] (te7) {$f_{c} \leq f_{c_{max}}$};
  \node [bloque1, below of=te7, node distance=2.2cm] (te8) {End};
  \path [line] (te1) -- (te2);
  \path [line] (te2) -- (te3);
  \path [line] (te3) -- (te4);
  \path [line] (te4) -- node {YES}(te5);
  \path [line] (te5) -- (te6);
  \path [line] (te6) -- (te7);
  \path [line] (te4) -- node {NO} ++(1.5,0) -- +(0,-2.65) -- (te6);
  \path [line] (te7) -- node {YES} ++(-1.5,0) -- +(0,5.55) -- (te3);
  \path [line] (te7) -- node {NO}(te8);
\end{tikzpicture}
  \caption{Proposed methodology to assess nonlinearity in non-stationary time series.}\label{fig2a}
\end{figure}

It is widely accepted that most biomedical systems are dynamic and
produce nonstationary signals \cite{Rangayyan}; the presence of slow
varying trends is only one type of nonstationarities present in
physiological signals. So, the novelty of the present document is to
propose a methodology based on the TFTS data method (which from now
on will be called band-phase-randomized surrogate data method) that
allows us to assess nonlinearity in data with different kinds of
nonstationarities (e.g., spikes, abrupt changes in the dynamical
behavior). The proposed procedure is depicted in Fig \ref{fig2a}.

\subsubsection{Band-Phase-Randomized Surrogates}
Band-phase-randomized surrogate data method is, as mentioned, a
modification of the RP algorithm in which not all phases but a
portion of the phases in the high-frequency band are
randomized. \\
Unfortunately, as stated by \cite{Theiler1993} it is difficult to
automate the procedure in order to make it applicable to all time
series. The methodology proposed in \cite{tomomichi2} to find de
correct value of $f_{c}$ (i.e., the correct portion in the frequency
band in which the phases are to be randomized) is only useful when
data have a slow varying trend, because when this statement is not
true, the stoping criterium is never met (i.e., $\mathcal{AC}(\tau =
1$) of data falls outside the distribution of surrogates ) and so
one always ends up using the iAAFT algorithm even when data is
not-stationary. In \cite{Guarin2010}, we propose that the stoping
criterium should be the similarity between data and surrogates,
i.e., surrogates should preserve the local behavior of the data.
But, when the data is in fact nonlinear this criterium fails. Next,
we present a new method for selecting the correct parameter of the
algorithm.

It should be noted that the use of the end-phase-randomized
surrogate data method will not improve the type II error because if
the method fails to reject the null when all phases are randomized
(using some statistic) then it certainly will not be able to reject
the null when just a portion of the phases are randomized. On the
other hand, the type I error will be improved by means of this
method.
\begin{figure}[b]
\centering
  \includegraphics[width=9cm]{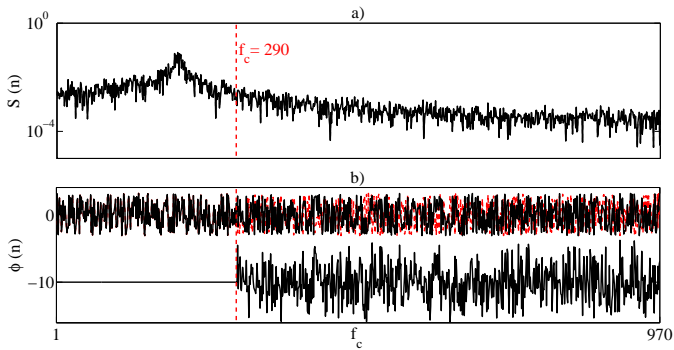}\\
  \caption{a) FT magnitude (note the logarithmic scale) and b) FT phases  as a function of $n$ for LS signal with $N=1940$ (continuos line) and one Band-Phase-Randomize surrogate $f_{c}=291$ (dotted line). $S(n)$ for data and surrogates are equal
for all $n$, but $\phi(n)$ is equal only for $n \leq f_{c}$. In this
case we are randomizing 70$\%$ of the higher frequency domain. In b)
the difference between the FT phases of data and surrogates is
displaced form cero for clarity.}\label{fig3}
\end{figure}
\subsubsection{Parameter selection}
To overcome the parameter selection problem we propose not to use
just one value of $f_{c}$ but a set of values. The proposed
methodology is as follows: First, we select two values $f_{c_{max}}
\approx N/2$ and $f_{c_{min}}$. Within this range, we select a set
of values for $f_{c}$ (e.g., 10 values), then we generate
Band-Phase-Randomized Surrogates using all those values and finally
we perform the nonlinearity test (one must ensure that linear
correlations of the data are preserved in surrogates for those values of $f_{c}$).\\
There are several ways to determine the value $f_{c_{min}}$; if the
Fourier transform magnitude ($S(n)$) has a pronounced peak then,
$f_{c_{min}}$ is selected above the peak (see Fig. \ref{fig3} a)).
If $S(n)$ does not have a pronounced peak (or has several) then
$f_{c_{min}}$ should be selected as the lowest value for which the
local mean of the data is preserved in the surrogates (see Fig.
\ref{fig5a}); when data have a pronounced peak, both criteria result
in a similar value of $f_{c_{min}}$.

\subsection{Selection of the discriminant statistic}
Dynamical measures  are  often  used  as  discriminating statistics,
the correlation being dimension one of the most popular choices
\cite{small6}. To estimate these, we first need to reconstruct the
underlying attractor. For this purpose, a time-delay embedding
reconstruction is usually applied. But this method is not useful for
data exhibiting nonstationarities because at the moment, there is
no optimal method for embedding such data \cite{bookthree}.\\
Therefore, as discriminant statistics we chose the Average Mutual
Information ($\mathcal{I}(\tau)$) \cite{bookthree}. The
$\mathcal{I}(\tau)$ is a nonlinear version of the
$\mathcal{AC}(\tau)$. It can answer the following question: On
average, how much does one learn about the future from the past? So,
we expect that if our data is not just a realization of a linear
non-stationary noisy process it would have a larger
$\mathcal{I}(\tau)$ than that of the surrogates.
\subsection{Implementation} Prior to the application of the methodology, we normalize
the data to zero mean and unit variance and find the largest
sub-segment that minimizes the end-point mismatch (this step is
extremely important and can be done automatically as suggested in
\cite{Schreiber2000}); if the data have a trend then one can
apply the preprocessing methodology proposed in \cite{tomomichi2}.\\
In order to reject a null hypothesis we generate $M=99$ surrogates
using an improved Amplitude Adjusted version of the
band-phase-randomized surrogate data method, because as the
$\mathcal{I}(\tau)$ depends on the data $\mathcal{AD}$, we have to
generate surrogates with equal $\mathcal{AD}$ as the data to avoid
false rejections. Then we compute the $\mathcal{I}(\tau = 1)$ for
the ensemble of surrogates and for the original time series (in a
previous study we showed that $\mathcal{I}(\tau)$ is sensible to the
type of dynamics only for small lags \cite{Guarin}). If
$\mathcal{I}(\tau = 1)$ is greater than that of the surrogates we
reject the null hypothesis at the 0.01 significance level;
otherwise, we do not reject the null.
\section{Results}
\subsection{Numerical results}
\begin{figure}
  \includegraphics[width=9cm]{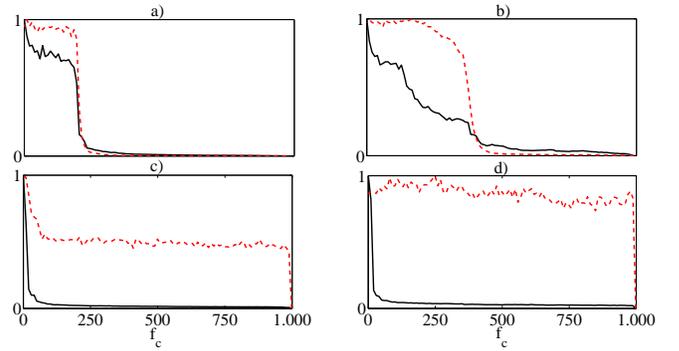}\\
  \caption{Normalized rms difference between local mean (continuos line) and variance (dotted line) of data (a) LS signal, b) LNS signal, c) NLS signal and d) NLNS signal) and Band-Phase-Randomize surrogates as a function of
$f_{c}$. The local mean and variance was calculated using windows of
length 64 with 50$\%$ overlap.}\label{fig5a}
\end{figure}
\begin{figure*}[t!9]
\centering
  \includegraphics[width=15cm]{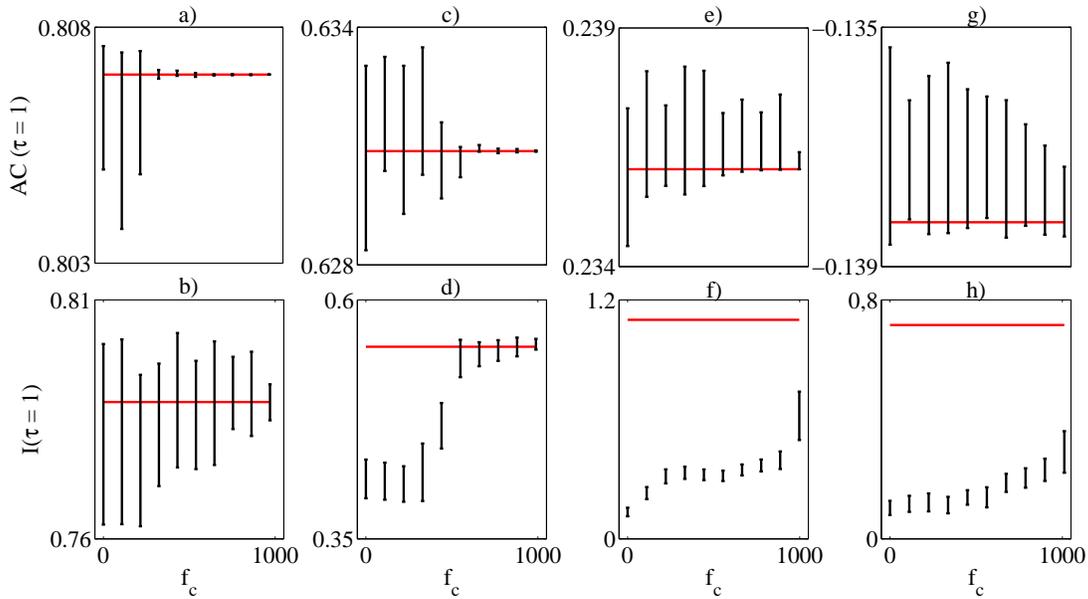}\\
  \caption{(Color online) a), c), e) and g) $\mathcal{AC}(\tau = 1$)of the original time series (a),b) LS signal, c),d) LNS signal, e),f) NLS signal and g),h) NLNS signal) (continuos vertical line)
and $\mathcal{AC}(\tau = 1$) of an ensemble Band-Phase-Randomize
surrogates (5th, 50th and 95th percentiles) as a function of $f_{c}$
($f_{c}=0$ is the result of using the iAAFT algorithm). b), d), f)
and h) the same as above but using the $\mathcal{I}(\tau =
1)$.}\label{fig6}
\end{figure*}

Prior to testing for nonlinearity we normalized each time series to
zero mean and unit variance, and selected a subsegment of the
signals that minimized the end-point mismatch, we end up with N=
1940, 1954, 1996 and 2023 number of data points for each time
series.\\ Fig \ref{fig5a} shows the normalized root mean square
(rms) difference between data (a) LS signal, b) LNS signal, c) NLS
signal and d) NLNS signal) and Band-Phase-Randomized surrogates as a
function of $f_{c}$ (when $f_{c}=0$ Band-Phase-Randomized surrogates
and the iAAFT surrogates are equivalent). It can be noted that for
linear data it is possible to obtain surrogates with almost the same
local behavior as the original time series while for nonlinear data
the local variance of surrogates is never similar to the data
(except for $f_{c}=N/2$). This result is expected because the
variance is a nonlinear statistic and surrogates are only
constrained to sample mean, sample variance $\mathcal{AD}$ and
$\mathcal{AC}(\tau)$ of data. \\From Fig. \ref{fig5a} we notice that
$f_{c_{min}} \approx$ 280, 400, 50 and 50 for each time series.
Anyhow, we use $f_{c_{min}} = 0$ and
$f_{c_{max}}=N/2-10$  for the following result. \\
Fig. \ref{fig6} shows the $\mathcal{AC}(\tau = 1$) and
$\mathcal{I}(\tau = 1)$ from data and Band-Phase-Randomized
surrogates. It can be noted that for linear stationary data (fig.
\ref{fig6} a) and b) ) the hypothesis tested by the iAAFT algorithm
cannot be rejected ($f_{c}=0$) and as expected, randomizing only a
portion of the higher frequency domain, does not affect this result.
When data is nonlinear (stationary or not) the test rejects the null
hypothesis of linearity for all values of $f_{c}$ within the
selected range of values. As shown in fig. \ref{fig6} g),
$\mathcal{AC}(\tau = 1)$ of data is not similar to that of
surrogates for some values of $f_{c}$, this implies that linear
correlations of the data are not well preserved in the surrogates
and one should not perform the nonlinearity test for these values of
$f_{c}$. In spite of this, the hypothesis is rejected.\\ The most
interesting case (at least for the purpose of the present document)
is the linear non-stationary case; in this situation nonlinearity is
detected using the iAAFT algorithm (fig. \ref{fig6} d), $f_{c}=0$),
so a careless application of the linear surrogate data method would
result in a false detection of nonlinearity (type I error). But, as
shown in fig. \ref{fig6} d), the nonlinearity is detected only for
certain values of $f_{c}$, in this case when $f_{c}>500$
nonlinearity is no longer detected by the test (the same curve as
fig. \ref{fig6} d), is obtained when the value of $M_{T}$ in
(\ref{ecu2}) is slightly modified, the range of
values of $f_{c}$ for which the null is rejected vary with $M_{T}$). \\
Two other important aspects can be noticed in Fig. \ref{fig6},
first, it is remarkable that when local mean and variance of
surrogates are similar to data, $\mathcal{AC}(\tau = 1)$ of ensemble
of surrogates is almost the same as data, this can be seen in Fig.
\ref{fig6} a) and c) for $f_{c}>300$ and $f_{c}>500$ respectively
(compare this with the results shown in Fig. \ref{fig5a} a) and b)),
but the same results are not observed when local variance of
surrogates is not similar to data (although the local mean of
surrogates is similar to data), this can be seen in Fig. \ref{fig6}
e) and g) respectively (compare this with the results shown in fig.
\ref{fig5a} c) and d)). Second, besides differentiating between
linear and nonlinear time series (stationary or not), this test can
be used to discriminate between linear stationary and linear
non-stationary data, in the former case the hypothesis of linearity
will be accepted for all values of $f_{c}$, while in the latter this
will occur only for certain values of $f_{c}$ (as shown in Fig.
\ref{fig6} d)).

To test the robustness of the method we performed the same analysis
presented here adding a 5$dB$ white noise to each time series and
found similar results.
\subsection{Application to HRV signals}
\begin{figure}[b!]
  \includegraphics[width=9cm]{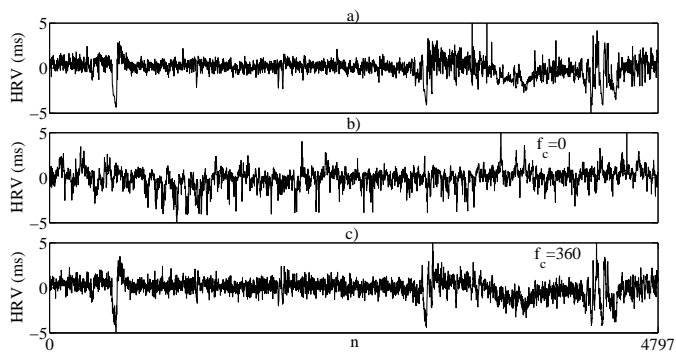}\\
  \caption{a) Segment of a HRV time series of a 32 year old healthy male, b) surrogate generated using the iAAFT
algorithm, c) band-phase-randomized surrogates using
$f_{c}=360$.}\label{fig9}
\end{figure}

Despite the fact that nonlinear dynamics are involved in the genesis
of HRV as a result of the interactions among hemodynamic,
electrophysiological, and humoral variables \cite{taskforce1996},
there is no proof that the recorded HRV time series (usually derived
from an ECG) reflects this nonlinearity, this must  be proven in
each case. In this section, we apply the proposed methodology to
assess nonlinearity in HRV which are known to have spikes and
nonstationarities due to variation of the patient
activity (see Fig. \ref{fig9} a).\\
Fig. \ref{fig9} a), shows a 1 hour record of the HRV of a healthy 32
year old male, the starting time is about midnight and the patient
is at rest. Fig \ref{fig9} b), depicted one surrogate time series
generated using the classical method (iAAFT surrogates), while
surrogates presented in Fig \ref{fig9} c), where generated using the
band-phase-randomized surrogate data method with $f_{c}=360$. \\ The
original time series has much of its energy concentrated in the high
frequency components, and as in the iAAFT surrogates the high
frequency energy of the original time series is blurred in all the
frequency spectrum, one gets surrogates that are not simular to the
HRV signal, allowing a trivial rejection of the null hypothesis.
Band-phase-randomized surrogates overcome this problem by preserving
the phases in a portion of the frequency spectrum, in this way, high
frequency and low frequency components of the original time series
are preserved in surrogates, as can be seen in Fig. \ref{fig9} a)
and c).

%

\begin{figure}[t!]
\centering
  \includegraphics[width=9cm]{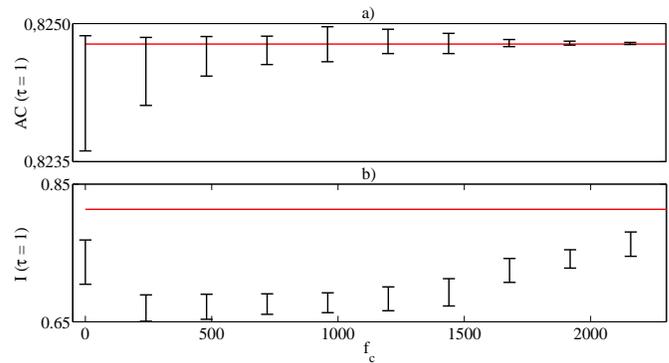}\\
  \caption{a) $\mathcal{AC}(\tau = 1$), b)$\mathcal{I}(\tau = 1)$ for the HRV signal and Band-Phase-Randomize surrogates
  as a function of $f_{c}$.}\label{fig10}
\end{figure}

Using the proposed methodology it was found that $f_{c_{min}}=200$
and $f_{c_{max}}=2300$, with this information, Fig. \ref{fig10} was
generated.

As expected, the null tested by the iAAFT surrogates is rejected
($f_c=0$), but as seen in Fig. \ref{fig6} d), this is not an
indicator of nonlinearity, but of nonlinearity or nonstationarity,
and as in this case it is acknowledge that the tested signal is
nonstatioanary, this test is not giving any new information about
this signal. But the proposed methodology is; it can be noticed that
when $f_{c}$ is within the selected range, the null hypothesis is
always rejected (and the linear correlations of the original time
series are always preserved in surrogates), and as was already
noticed (Fig. \ref{fig6} f) and h)), this is a clear indicator of
the presence of nonlinear correlations. By this means, we confirm
that there is a complex nonlinear physiological process underlying
the HRV.
\section{Conclusion}
In this document, we presented a methodology based on the TFTS data
method and the iAAFT algorithm that allows us assess nonlinearity in
non-stationary time series. Based on some simulated data we
demonstrate that our methodology is able to differentiate between
linear stationary, linear non-stationary (even when the linear data
is transformed by a nonlinear monotonic static observation function)
and nonlinear time series. This method is different from previously
proposed nonlinearity tests because: i) we do not randomize the
phases in all the frequency domain but in a portion of the frequency
domain , and ii) we do not select a correct value of $f_{c}$ but a
correct range $[f_{c_{min}}, f_{c_{max}}]$, and within this range, a
set of values for the parameter $f_{c}$.\\
Applying this test to physiological time series, we found that
nonlinear correlations are present in HRV signals of a healthy male, this confirms that nonlinear dynamics are
involved in the genesis of HRV, but as mentioned, every times series should be tested because there no a priori method
to determine if a given signal represent the nonlinearity of the process.\\
It is worth mentioning that as pointed out by many authors
(\cite{tomomichi2}, \cite{Theiler1993}), the linear surrogate data
methods are only suitable for stochastic like data, and as the
present methodology is based on that, the same limitations apply.

\section*{Acknowledgment}

D. Guarín was supported by the UTP and COLCIENCIAS, program
\emph{Jóvenes Investigadores e innovadores 2010}. E. Delgado is
supported by the Condonable Credits program of COLCIENCIAS in
Colombia. Additionally, he would like to thank to the Research
Center of the ITM of Medellín – Colombia who supported him with the
PM10201 grant.


%

%


\ifCLASSOPTIONcaptionsoff
  \newpage
\fi



\bibliographystyle{IEEEtran}
\bibliography{bib}

\begin{thebibliography}{10}
\providecommand{\url}[1]{#1}
\csname url@samestyle\endcsname
\providecommand{\newblock}{\relax}
\providecommand{\bibinfo}[2]{#2}
\providecommand{\BIBentrySTDinterwordspacing}{\spaceskip=0pt\relax}
\providecommand{\BIBentryALTinterwordstretchfactor}{4}
\providecommand{\BIBentryALTinterwordspacing}{\spaceskip=\fontdimen2\font plus
\BIBentryALTinterwordstretchfactor\fontdimen3\font minus
  \fontdimen4\font\relax}
\providecommand{\BIBforeignlanguage}[2]{{%
\expandafter\ifx\csname l@#1\endcsname\relax
\typeout{** WARNING: IEEEtran.bst: No hyphenation pattern has been}%
\typeout{** loaded for the language `#1'. Using the pattern for}%
\typeout{** the default language instead.}%
\else
\language=\csname l@#1\endcsname
\fi
#2}}
\providecommand{\BIBdecl}{\relax}
\BIBdecl

\bibitem{theiler}
J.~Theiler, S.~Eubank, A.~Longtin, B.~Galdrikian, and J.~D. Farmer, ``Testing
  for nonlinearity in time series: The method of surrogate data,''
  \emph{Physica D}, vol.~58, p. 77 – 94, 1992.

\bibitem{theiler2}
J.~Theiler and D.~Prichard, ``Constrained-realization monte-carlo method for
  hypothesis testing,'' \emph{Physica D}, vol.~94, no.~4, pp. 221 -- 235, 1996.

\bibitem{Kaplan1997}
D.~T. Kaplan, \emph{Frontiers of blood pressure and heart rate analysis}.\hskip
  1em plus 0.5em minus 0.4em\relax Amsterdam: IOS Press, 1997, vol.~35, ch.
  Nonlinearity and Nonstationarity : The use of surrogate data in interpreting
  fluctuations, pp. 15 -- 28.

\bibitem{Schreiber2000}
T.~Schreiber and A.~Schmitz, ``Surrogate time series,'' \emph{Physica D}, vol.
  142, pp. 346--382, 2000.

\bibitem{Timmer1998}
J.~Timmer, ``Power of surrogate data testing with respect to nonstationarity,''
  \emph{Physical Review E}, vol.~58, no.~4, pp. 5153 -- 5156, 1998.

\bibitem{Schreiber1998}
T.~Schreiber, ``Constrained randomization of time series data,'' \emph{Physical
  Review Letters}, vol.~80, no.~10, pp. 2015 -- 2018, 1998.

\bibitem{Schmitz99surrogatedata}
A.~Schmitz and T.~Schreiber, ``Surrogate data for non-stationary signals,'' in
  \emph{Workshop on Chaos in Brain?}, K.~Lehnertz, J.~Arnhold, P.~Grassberger,
  and C.~E. Elger, Eds.\hskip 1em plus 0.5em minus 0.4em\relax Singapore: World
  Scientific, 1999, p. 222–225.

\bibitem{tomomichi}
T.~Nakamura and M.~Small, ``Small-shuffle surrogate data: Testing for dynamics
  in fluctuating data with trends,'' \emph{Physical Review E}, vol.~72, pp.
  056\,216--1 -- 056\,216--6, 2005.

\bibitem{tomomichi2}
T.~Nakamura, M.~Small, and Y.~Hirata, ``Testing for nonlinearity in irregular
  fluctuations with long-term trends,'' \emph{Physical Review E}, vol.~74, pp.
  026\,205--1 -- 026\,205--8, 2006.

\bibitem{Theiler1993}
J.~Theiler, P.~S. Linsay, and D.~M. Rubin, ``Detecting nonlinearity in data
  with long coherence times,'' in \emph{Predicting the future and understanding
  the past}.\hskip 1em plus 0.5em minus 0.4em\relax Addison-Wesley, 1993, pp.
  429--455.

\bibitem{Faes2009-2}
L.~Faes, H.~Zhao, K.~H. Chon, and G.~Nollo, ``Time-varying surrogate data to
  assess nonlinearity in nonstationary time series: Application to heart rate
  variability,'' \emph{IEEE Trans. on Biomedical Engineering}, vol.~56, no.~3,
  pp. 685 -- 695, 2009.

\bibitem{Keylock2007}
C.~J. Keylock, ``A wavelet-based method for surrogate data generation,''
  \emph{Physica D}, vol. 225, p. 219–228, 2007.

\bibitem{Guarin2010}
D.~Guarín, A.~Orozco, E.~Delgado, and E.~Guijarro, ``On detecting determinism
  and nonlinearity in microelectrode recording signals: Approach based on
  non-stationary surrogate data methods,'' in \emph{32nd Annual International
  Conference of the IEEE EMBS}, 2010, pp. 4032 -- 4035.

\bibitem{Schreiber1996}
T.~Schreiber and A.~Schmitz, ``Improved surrogate data for nonlinearity
  tests,'' \emph{Physical Review Letters}, vol.~77, p. 635 – 638, 1996.

\bibitem{Venema2006}
V.~Venema, F.~Ament, and C.~Simmer, ``A stochastic iterative amplitude adjusted
  fourier transform algorithm with improved accuracy,'' \emph{Nonlinear
  Processes Geophysics}, vol.~13, p. 321–328, 2006.

\bibitem{small6}
M.~Small, T.~Nakamura, and X.~Luo, \emph{Nonlinear Phenomena Research
  Perspectives}.\hskip 1em plus 0.5em minus 0.4em\relax Nova Science
  Publications, 2007, ch. Surrogate data methods for data that isn't linear
  noise, pp. 55 -- 81.

\bibitem{Faes2009-1}
L.~Faes, H.~Zhao, K.~H. Chon, and G.~Nollo, ``A method for the time-varying
  nonlinear prediction of complex nonstationary biomedical signals,''
  \emph{IEEE Trans. on Biomedical Engineering}, vol.~56, no.~2, pp. 205 -- 209,
  2009.

\bibitem{Glass2009}
L.~Glass, ``Introduction to controversial topics in nonlinear science: Is the
  normal heart rate chaotic?'' \emph{Chaos}, vol.~19, pp. 028\,501--1 --
  028\,501--4, 2009.

\bibitem{Goldberger2000}
A.~L. Goldberger, L.~A.~N. Amaral, J.~M.~H. L.~Glass, P.~C. Ivanov, R.~G. Mark,
  J.~E. Mietus, G.~B. Moody, C.-K. Peng, and H.~E. Stanley, ``Physiobank,
  physiotoolkit, and physionet : Components of a new research resource for
  complex physiologic signals,'' \emph{Circulation}, vol. 101, pp. 215 -- 220,
  2000.

\bibitem{Rangayyan}
R.~M. Rangayyan, \emph{Biomedica Signal Analysis}, M.~Akay, Ed.\hskip 1em plus
  0.5em minus 0.4em\relax IEEE Press, 2002.

\bibitem{bookthree}
M.~Small, \emph{Applied Nonlinear Time Series Analysis - Applications in
  Physics, Physiology and Finance}.\hskip 1em plus 0.5em minus 0.4em\relax
  World Scientific, 2005.

\bibitem{Guarin}
D.~Guarín and A.~Orozoco, ``Pruebas de no linealidad para series temporales,''
  \emph{Submited}, 2010.

\bibitem{taskforce1996}
T.~F. of~The European Society~of Cardiology, T.~N. A.~S. of~Pacing, and
  Electrophysiology, ``Heart rate variability. standards of measurement,
  physiological interpretation, and clinical use,'' \emph{European Heart
  Journal}, vol.~17, pp. 354 -- 381, 1996.

\end{thebibliography}
\balance
\end{document}